\documentstyle{article}

\begin{document}

{\bf CLOCK AND CATEGORY: IS QUANTUM GRAVITY ALGEBRAIC? }

\bigskip

by Louis Crane, math department, KSU.

\bigskip

ABSTRACT: We investigate the possibility that the quantum theory of
gravity could be constructed discretely using algebraic methods. The
algebraic tools are similar to ones used in constructing Topological
Quantum Field theories. The algebraic structures are related to ideas
about the reinterpretation of quantum mechanics in a general
relativistic context.

\bigskip

I. INTRODUCTION

\bigskip

The histories of mathematics and theoretical physics are so intimately
interwoven that neither can really be understood in isolation from
the other. There is no example of a fundamental advance in theoretical
physics which did not involve a change in the mathematical structure
in which the physical theory is formulated.  On the other hand, a new
construction in mathematics very often
represents a distillation of a physical concept. Differential
four dimensional manifolds with lorentzian metrics, for example, can be thought
of as structures within which
Einstein's falling elevators can coexist. Turning points in the
development of theoretical physics are moments when mathematical
thought and fundamental conceptual physical thought achieve a sort of
fusion.

If we examine the current state of fundamental theoretical physics
from a mathematical point of view, we find an extremely odd situation.
The subject is dominated by the application of a formal
pseudomathematical device, the Feynman path integral, which has no
rigorous mathematical formulation, and which clearly cannot be given
one in any straightforward way, since most of the quantum theories
constructed from them do not actually exist. The disharmony between the
foundations
of mathematics and physics is so extreme and so jarring that most
workers in the field tend to blot it out of their minds.

Topological quantum field theories (TQFTs), were originally defined
formally in terms of path integrals [1,2]. Subsequently, they were
rigorously constructed in low dimensions by various authors, using
algebraic tools [3,4,5]. Among the algebraic tools which appear in this
process, the techniques of category theory play a special role.

The purpose of this paper is to propose that the same mathematical
tools which put TQFT on a rigorous footing can be applied to the
problem of constructing the quantum theory of gravity. The physical
thought at the heart of this proposal is the idea that path integral
technology is inappropriate for quantum gravity (or quantum gravity
coupled to matter) because the geometry which needs to be quantized
has a cutoff at the Planck scale. Thus, averaging over continuum
geometries via a path integral introduces unphysical infinities, which
destroy the mathematical structure of the theory.

One family of algebraic constructions of TQFTs uses
topological state sums. These can be thought of as discretized path
integrals which avoid the usual pitfalls of lattice theories by a sort
of algebraic magic.

What will be argued here is that the categorical tools which have been
applied to TQFT can be interpreted as producing quantum gravity as well.

Perhaps it is not very useful to try to argue the plausibility of such a
program in advance. It can really be substantiated
only by empirical success, which still lies in the future.
Nevertheless, let me make a few general remarks.
It is hardly a controversial observation that the quantization of
gravity is a deep problem. Hence it is reasonable to suggest that the
formulation of the quantum theory of gravity requires a new
mathematical structure to contain it. The insistence that fundamental
physics should be formulated in well defined, non-divergent terms seems
almost eccentric in the context of contemporary
theoretical physics, where the dramatic success of Feynmanology
has created a certain mind set over the last half century. I cannot
help feeling though that anyone who pauses to consider the matter would
find a rigorous mathematical foundation for a truly fundamental
physical theory highly desirable.

Let me also remark that categorical algebra and relativity theory have
similar philosophical roots. Both have in their foundations the idea
that things should be defined independently of a coordinate
system/observer.
The reinterpretation of quantum mechanics which will be explored below
is that the probability interpretation must be relativized, and that
the Hilbert spaces related to different observers have natural linear
maps between them. Thus, the observers form a category, which we can
construct, and which has a very special structure. The invariant
language of categories provides a form for discussing the relativity
of the observation process and probability interpretation with
respect to the observer.

The purpose of this paper is twofold: first to describe a new branch of
mathematics, namely the algebraic approach to topological quantum
field theory (TQFT) in three and four dimensions, and the transition
between the two by the picture called the dimensional ladder.
Secondly, to propose a line of reasoning by which
this mathematical structure could be used to produce a quantum theory
of gravity. The two parts are not at an equal level of development.
The algebraic constructions are understood at least in outline, while
the physical interpretation revolves around a reconsideration of the
role of the observer in quantum mechanics, and is still unclear.
Nevertheless, the resonances between the mathematical structure
presented here and the development of the Ashtekar/loop variables
approach to quantum general relativity [6] are so strong that the
suggestion that they are related deserves to be pursued.

In a certain sense, I am proposing that a particular three dimensional
TQFT, namely the CSW theory, already contains the quantum theory of
gravity, but that in order to introduce clocks into the theory and connect it
with experiment, it is
necessary to relate it to a four dimensional theory.

Accordingly, the structure of this paper is as follows. We first
review the structure of CSW theory in 3D, and explain its relationship
to quantum spin networks. Next, we explain how this can be thought of
as a quantum theory of gravity, assuming a suitable reformulation of
the principles of quantum mechanics. Then we shall discuss how to
relate 3D and 4D TQFTs, and last, discuss how this may be relevant to
the problem of introducing clocks into the theory.

The proposal discussed here creates a rather difficult communication
problem, since vanishingly few theoretical physicists study category
theory. An ambitious reader might want to study the subject from
Maclane [7]. It may also be possible to wing it and just follow my
development as it stands. Although it was clearly impossible to make
it self contained, I have interspersed my discussion with thumbnail
definitions which may suffice. Perhaps a physically minded reader would prefer
to
skip to chapter III on a first reading, where the physical ideas are
discussed.

\bigskip

II. THE BASIC STRUCTURE OF 3D TQFT

\bigskip

A. TQFT

\bigskip

The definition of a TQFT can be explained in many ways. A major theme
of this paper is that fundamental physical ideas related to quantum
gravity can best be understood in the language of categorical algebra.
Since this area is rather strange to physicists (and by no means
universally popular among mathematicians), let us begin by showing how
TQFTs, which, after all, are a type of physical theory, have a natural
description in categorical language.

Remember that a category has objects and morphisms between the
objects, which can be composed when the range of the first is the
domain of the second. The classical example is sets and functions.

Accordingly, let us begin by recalling the structure of a cobordism category.
The
category of oriented n-dimensional cobordisms has oriented compact n-1
dimensional manifolds as objects and cobordisms as morphisms. (A
cobordism from M to N is an oriented n-dimensional manifold P with boundary,
together with an oriented diffeomorphism between the boundary of P and
$M^* \bigcup N$.)
Composition of morphisms comes from gluing of manifolds along shared
boundary components.

The category of oriented n-cobordisms has the natural structure of a
tensor category with duality. (A tensor category is one where two
objects can be combined into a third by an operation we call tensor
product, and where morphisms can be similarly combined. The prototype
example is the category of vector spaces.)
The tensor product is disjoint union and the
duality is reversal of orientation.

The most elegant definition of an n-dimensional TQFT is that it is a
monoidal
functor from the category of oriented n-cobordisms with disjoint union
as tensor product to the category
{\bf VECT} of
finite dimensional vector spaces with the usual tensor product i.e. a
functor which preserves tensor product up to canonical coherent
isomorphism.( It is a point often missed that this suffices---that
a manifold with opposite orientation is sent to the dual space of the
image with the given orientation is an easy theorem, not a necessary
part of the definition).
Let me remind the reader that a functor is a mapping from one category
to another, taking objects to objects and morphisms to morphisms, and
taking composition to composition.

We often modify the definition of TQFTs by modifying the cobordism
category. For instance, we can specify a framing of the tangent
bundle of the cobordisms and of a formal neighborhood of the closed
manifolds.  Another possibility is to include insertions of submanifolds
 in the
manifolds and matching insertions in the cobordisms. We also refer to
tensor and duality preserving functors from such modified cobordism
categories to {\bf VECT} as TQFTs.

Let us spell out this definition for the less categorically inclined.
An n-dimensional
TQFT assigns a vector space to each oriented n-dimensional manifold,
and a
linear map to each oriented cobordism in such a way that the
composition of cobordisms corresponds to the composition of linear
maps, the disjoint union of manifolds gets the tensor product of
vector spaces, and the manifold with opposite orientation gets
assigned the
dual space. As John Baez has pointed out [8], there is a natural inner
product on the vector spaces whenever they are associated to a
bounding 3-manifold, generated by gluing pairs of links along the
boundary. Thus, we can think of them as Hilbert spaces.

Thus, for n=3, a TQFT assigns a vector space to a surface, and a
linear map to a 3-dimensional cobordism. We also think of an
assignment of a vector space to a surface with labelled punctures and
a linear map to a 3D cobordism which contains a link with components
which can end on the punctures, with labels on the components and
framings of the cobordism and link as a 3D-TQFT. CSW theory in fact
provides us with a 3D-TQFT in this extended sense. The similarity
between this and the loop variable picture for quantum gravity [6] should
be evident to the initiated.

\bigskip

B THE CSW TQFT

\bigskip

The 3D TQFT which seems to be related to the quantum theory of gravity
was first constructed in physicist's language by Witten from a
topologically invariant lagrangian by means of a path integral [1].
Subsequently, it was reconstructed by algebraic methods by a number of
authors. We will be interested in two approaches to the algebraic
construction of 3D-TQFTs: the construction from a modular tensor
category in [5], and the topological state sum, also from a modular
tensor category in [4], but since generalized to any tensor category
with only a weak condition on its duals in [9]. (See [10] for the
definition of a modular tensor category.) The construction in [3] is
also elegant and historically important, but does not play as big of a
role in what follows.

The point which is most important for us here is the form of the
construction of the finite dimensional Hilbert spaces on the surfaces,
possibly with labelled punctures, from the modular tensor category.

As originally explained in [10], the vector space attached to a surface
with labelled punctures is constructed by translating the surface into
a word in the tensor category. This is accomplished by cutting the
surface up into three holed spheres, or trinions, and attaching a
tensor product to each trinion. The axioms of a modular
tensor category are just what is needed so that if the same surface is
decomposed in two different ways, an isomorphic object results.
In order to assign a vector space to the punctured surface, we extract
the invariant part of the object (this is the set of homomorphisms of
the identity object of the category into the object; hom sets in these
categories form vector spaces.)

In the case of the modular tensor category associated to SU(2) at any
level, we can write a basis for the vector space attached to the
surface by writing all ways to label the cuts with irreducible objects
in the category, consistent with the quantum version of the
Clebsch-Gordon relations. (A modular tensor category possesses only a finite
set of irreducibles.) The formula for a general modular tensor
category uses combinations of the spaces of tensor operators on
irreducibles. We will not need it here, see [10].
Two different decompositions of the surface give two different bases,
and the transition coefficients between them are determined by the
structure of the category.

The example which seems relevant to quantizing gravity is the SU(2)
case, ie the category either of representations of the loop group on
SU(2) at a central extension, or of the quantum group at a root of
unity with dimension zero modules quotiented out. The simple form of
the vector spaces in that case seems to be closely related to a
geometric interpretation. The thing which needs observing at this
stage is that the irreducible objects of the category are labelled by
spins, much like the irreducible representations of SU(2). It may turn
out that natural models for gravity coupled to matter emerge from more
complex modular tensor categories.

In summary, the vector spaces attached to a punctured surface have
bases described by cutting the surfaces into trinions (with the
punctures, labelled with spins, at open ends of trinions) and giving
labels to the internal cuts from the set of allowable spins, so that
the triple of spins on each trinion satisfies the Clebsch-Gordon
conditions for the truncated category of representations of a quantum
group at a root of
unity.

We now need to know a little about how to extract numbers and linear
maps from 3D manifolds in this theory. There are two simple
observations which solve this problem. One is that only the trivial
representation can be extended across a closed disk. This means that
the vector in the vector space associated to a compact surface which
is associated to a solid handlebody which bounds the surface is the
vector in the basis corresponding to any trinion decomposition which
extends to the solid handlebody in which all the cuts are labelled by
the trivial representation.

The second observation is that the change of basis from one trinion
decomposition to another allows us to obtain a linear representation
of the group of large diffeomorphisms (diffeomorphisms modulo isotopy)
of the surface. Thus, a vector in the basis labelled by some spins on
the cuts of a decomposition is mapped into some linear combination of
similar vectors by a map of the surface to itself, where the
coefficients are determined by the structure of the category.

When these two ideas are combined, we can derive formulas for the
invariants on closed manifolds, and the linear maps on manifolds with
boundary. The long and short of it is that everything can be written
as linear combinations of the numbers we get by taking the dual
pairing of a state on a surface embedded in $R^3$ given by labelling
the cuts of the trinion decomposition by spins with the state obtained
by labelling the cuts of any trinion decomposition which extends to
the exterior with the trivial representation. (The second state is the
"exterior vacuum").

We can shrink the labelled embedded surface until it is very thin, and
think of the number we are obtaining as an invariant of a labelled
knotted trivalent graph. This is a generalization of the Jones
invariant to knotted graphs. (The generalisation to an arbitrary
modular tensor category requires us to label the edges of the embedded
graph
with
irreducible objects of the category and the vertices with tensor operators).

\bigskip

C. CSW THEORY AND DEFORMED SPIN NETWORKS

\bigskip

The invariants of embedded labelled graphs at the heart of the CSW
TQFT are closely related to the evaluations of spin networks due to
Penrose [11]. Spin networks are not sensitive to embeddings, but that is
due precisely to the fact that Penrose did not have quantum groups to
work with, and used representations of ordinary SU(2) instead [12].

Spin networks have been undeservedly forgotten. They were an effort to
produce the kind of quantum geometry necessary to quantize gravity,
and they succeeded to the point of giving a plausible form of
quantized 2+1 gravity. One point of view on the present work is that
it is an effort to bring Penrose's program up one dimension by adding
better algebra.

The best way to understand how spin networks apply to quantizing
geometry is to follow the approach of Regge and Ponzano [13].
What they do is to triangulate a region of space surrounded by the
embedded graph, and prove by an inductive argument that the evaluation
of the graph is given by a formula involving a sum over labellings of
a product of one 6J symbol for each tetrahedron.

In [14], it was observed that this formula is equally valid for the CSW
invariants of embedded graphs. Let us note that this formula, using
the representations of a quantum group at a root of unity, is
identical with the formula used to construct a 3D-TQFT in [5]. It is
being applied here in a different way to obtain the CSW invariant
instead. Specifically, it is being applied to a triangulation of one
half of the manifold only. The Viro-Turaev formula gives the absolute
square of the CSW invariant by summing over the entire manifold. (It
would be very attractive to find a new formula of a similar form which
bridged this gap, ie gave CSW when computed on the entire manifold.
Kuperberg [15] has given a candidate for such a formula, which
reproduces CSW for embedded graphs, but then gives zero on most
manifolds. It is not yet clear if his work admits a modification to
reproduce CSW on the nose.)

This formula is the classical example of what we mean by a topological
state sum. It is a summation on a triangulation which does not change
if we alter the triangulation. It has the form of a lattice field
theory, with a strange lattice-independence which comes from abstract
algebra.
(The basic topological invariance of our formula comes from the
Biedenharn-Eliot relation [16], as the chemists know it, which is the
Stasheff pentagon for the associator of the tensor category).

\bigskip

III. CSW THEORY AND QUANTUM GRAVITY

\bigskip

The above formula, in either context, has a natural interpretation as a
quantum geometry. We can reinterpret the spins as lengths. We are then
summing over all assignments of lengths to edges in a triangulation,
and we can interpret this as a discretized summation over metrics. If
we make this interpretation, and ask which terms in our sum have
stationary phase, we find that they are discrete analogs of flat
metrics. It was this observation which made Regge and Ponzano
interpret the formula, in the case of representations of ordinary
SU(2), as a discretized path integral for 2+1 dimensional gravity.
(The equations of motion for 2+1 general relativity are exactly
flatness).

Thus, if I were to argue that CSW theory gives us 2+1 gravity, the
argument would be quite straightforward. Indeed
, it would also not be terribly original. A number of authors have
made just such a connection in a number of ways [17]. Nobody expects 2+1
dimensional gravity to behave very differently from a TQFT.

Why then is the suggestion in this paper so novel?
The physical objection to the idea that 3+1 gravity has a similar
formulation to the 2+1 theory is based on the observation that the {\bf
linearized}
theory of 3+1 dimensional gravity contains excitations called
gravitons, which are similar to photons in that they come in all
frequencies. These can be combined into wave packets of arbitrarily
small volume, leading to the twin conclusions that quantum gravity A)
contains an infinite dimensional Hilbert space, and B) has local
excitations, both of which are properties alien to TQFT.

I want to argue that both of these arguments are misleading.
Linearized general relativity is an extremely poor model for the full
theory. It possesses a background metric, which solves the probability
interpretation problems, at the cost of destroying the  fully
invariant nature of the theory. In particular, both properties A and B
above are artifacts of the linearized theory, which do not survive in
the full theory. As regards property A, it is extremely implausible
that gravitons of wavelength below the Planck scale survive the
interactions of the full theory. They are
more likely to disappear into a cloud of Planck scale black holes
under their self gravitation. This puts an ultraviolet cutoff on the
local excitations, and makes the Hilbert space finite dimensional.
Recent works of several authors have pointed out that the energy of
states in any bounded region is is cut off at the Schwartzschild mass,
so the dimension of the space of states is finite, and related to the
surface area.

Far worse is the  idea that excitations in general relativity are
"local" in the same sense as in the linearized theory. General
relativity is diffeomorphism invariant. This means that a state
where an excitation occupies a given point (or, more plausibly, small
volume) on the manifold is gauge equivalent to a state in which it is
someplace else. This is well known to relativists, who say that "the
points disappear".

General relativity has two other properties of a similar type. One is
that the Hamiltonian vanishes, ie, that the time evolution operator is the
identity. (Let us observe that the identical fact holds for a TQFT).
Another is that the inner product on Hilbert space which provides a
measure for the probability interpretation is
missing, or at least well hidden. Various relativists have attempted
to find the inner product by elaborate measures, but without clear
success. It is, in any case, hard to imagine what a probability
interpretation for whole universes would mean, since there can be no
external observer.

These properties are so maddeningly counterintuitive that most
physicists contrive to forget them. I propose to approach the problem
of quantum gravity by recasting the form of the theory in such a way
as to make these properties central. My proposal was originally motivated more
by
the mathematical tools described above than by direct reflection on
the subtleties of quantum gravity. Nevertheless, I would like to state
it in physical terms before pointing out how the techniques of TQFT
could be used to construct the theory.

What I am proposing is that quantum gravity is not a quantum field
theory at all. Instead of a Hilbert space and operator fields there is
a much subtler structure, which I call a model for quantum gravity.
Here is an outline of the principles and corresponding structure of
such a model:

\bigskip

{\bf PRINCIPLES FOR A MODEL FOR THE QUANTUM THEORY OF GRAVITY}

\bigskip

1.{\it No observation is possible without an observer.} Hence there is no
Hilbert space associated with a closed universe. Any observer is part
of a universe, hence occupies a 3-manifold with boundary, and makes
observations on another such with a shared boundary.

2.{\it There is no observation at a distance.} Thus the Hilbert spaces in
the theory reflect the interface between observer and system. This
means they are associated to surfaces.

3.{\it Observers observe each other.} Hence there are linear maps between
the Hilbert spaces attached to surfaces in the same 3-manifold. These
maps satisfy a natural consistency condition.

4{\it .States for quantum gravity can be described by embedded graphs or
knots.} Hence the assignments of Hilbert spaces and maps above can be
extended to surfaces with punctures and 3-manifolds with graphs which
end on the punctures of their boundaries.

5. {\it General relativity is diffeomorphism invariant.} Hence all the above
assignments depend only on topological type.

6.{\it General relativity is a theory of geometry.} Hence a state in a
special basis for the
Hilbert space on a surface must assign values to lengths on the
surface, and probability amplitudes to combinations of lengths elsewhere.

7.{\it The Hamiltonian for general relativity is 0.} Hence there is no time
evolution for states in themselves. The core theory is three
dimensional. The experience of time evolution in the world must be
described by treating variables in the theory as local clocks. There
are no global clocks.

8.{\it A global Hilbert space can be recovered in a semiclassical
limit}. If we associate a family of observers in some natural way to a
triangulation of the manifold, we can use the maps between the vector
spaces to take an inverse limit. In the limit as the triangulation is
refined this reproduces a global Hilbert space, which can be thought
of as corresponding to a universe full of classical observers whose
quantum correlations can be neglected.

\bigskip

We invite the reader to contemplate the intrinsic plausibility of this
proposal before considering the following observation:

\bigskip

{\large THE CSW TQFT PROVIDES EXACTLY WHAT WE HAVE SPECIFIED ABOVE AS
A MODEL FOR QUANTUM GRAVITY, except for the local clocks.}

\bigskip

Let us spell this out: the vector spaces on embedded surfaces correspond to the
spaces which an
observer on one side can observe when interacting with the other. The
linear maps between these are the maps corresponding to the cobordisms
between them given by the regions of space they jointly bound. The
formula given above for CSW invariants as a summation is interpreted
as a probability amplitude for lengths in the volume they enclose.
Spins are reinterpreted as lengths, as explained above.

In some sense, then, we are suggesting that we already know the
quantum theory of gravity. The deformed spin network description of
states in CSW on a surface allows us to interpret them as quantum
3-geometries on regions thought of as the interiors of observed
systems.

In a theory with vanishing Hamiltonian, the state of the whole
universe cannot change. A model for the quantum theory of gravity is
the same as a state for the universe. Thus,
this proposal is consistent with the observation that the Chern
Simons lagrangian gives a state for quantum general relativity in the
Ashtekar variables, with cosmological constant [18]. This proposal is
also related to the suggestion of Witten that the Chern Simons theory
is the topological phase of gravity [19], together with the suggestion
that we never left the topological phase.

Of course, this is not what anybody really wants from a theory. In
order to relate to experiment, it is necessary to describe a time in
which things change.

The rest of my proposal is a line of development which attempts to
solve this problem by using an important fact about TQFTs, namely that
the algebraic constructions of TQFTs in adjacent dimensions are
closely related. The idea is that a 4D-TQFT which is closely related
to CSW theory would contain the secret of time.

\bigskip

IV. 4D TQFT AND THE DIMENSIONAL LADDER

\bigskip

The goal of the mathematical program which is outlined here is to
produce a 4D-TQFT which has three properties: it is factorizable,
related by categorification to
the 3D CSW theory, and admits a topological state sum formulation.

We shall explain here what each of these three condition means, and
suggest how they can be obtained. In the last part of this paper, we
shall try to show what each of them has to do with quantizing gravity,
(and incidently why categories are related
to clocks).
\bigskip

A. Factorizability
\bigskip

A TQFT can be thought of as an invariant of closed manifolds which can
be "factorized" when the manifold is cut along a hypersurface so that
the two halves (manifolds with boundary) are assigned vectors in a
vector space, and the invariant is recovered as a dual pairing.
A 3D-TQFT with factorizability has an analogous structure one layer
farther down in dimension so that we can cut surfaces along sets of
circles and write the vector space on the surface as the hom-space
between objects associated to the pieces in a suitable category.
(the categorical analogue of
a dual pairing!). The definition of a TQFT with factorizability is
similar in higher dimension, except that there are successive
categorical layers at higher codimension. See [20] for the general
definition. We shall give the 3D definition in its full glory in a
minute.

The importance of factorizability is that it allows us to obtain the
objects on manifolds with boundary by cutting them up into manifolds
with corners. In particular, a 3D theory can be extended to include
three manifolds with corners. If we have a 3-manifold with boundary
with an inserted link with ends on the boundary, we can describe this
as a 3 manifold with corners by hollowing out a neighborhood of the
link. The circles around the punctures at the ends of the links then
are the corners. This means that the extension of CSW theory to
include punctures and links or embedded graphs is a direct consequence
of factorizability. As we shall see, factorizability will be essential
in trying to make a physical interpretation of our theory.

For completeness, we reproduce here the definition from [20] in both a
categorical and not-so categorical form.

To set up the formal definition embodying this notion, we must
remind the reader that a finitely generated
 semisimple linear category is one in which each
object is isomorphic to a direct sum of irreducible (simple) objects chosen
from a finite set of such objects, hom-sets (the set of morphisms
between two objects) are complex
vector spaces, and composition is bilinear. As categories, they are equivalent
to ${\bf VECT}^n$ for some $n$. For the theory of such
categories, also called {\bf VECT} modules, see [21]. As shown in [21],
these categories form a monoidal bicategory:
objects are {\bf VECT} modules,
1-arrows are exact {\bf C}-bilinear functors, 2-arrows are natural
transformations, and the tensor product is given up to canonoical
equivalence by using pairs of the generating simple objects in the
tensorands as a set of generating simple objects.

Similarly observe that there is a monoidal bicategory of 3-dimensional
cobordisms with corners, $3-cobord_2$: its objects are 1-manifolds, its
1-arrows are (2-dimensional) cobordisms of 1-manifolds, and its
2-arrows are cobordisms with corners between pairs of 2-dimensional cobordisms
with the same source and target.
To be precise, a 3-dimensional cobordism with corners is a 3-manifold
with corners, whose boundary is a union along a family of circles
 joining corresponding boundary components
of the two surfaces. The 1-dimensional composition of 1-arrows and
2-dimensional composition of 2-arrows are just given by glueing target
to source. The 1-dimensional composition of 2-arrows consists of glueing
along the corners and glueing on a ``collar'' .
It is trivial to verify that disjoint union gives
this bicategory the structure of a monoidal bicategory.

In what follows, we shall refer to a cobordism (resp. cobordism with
corners) as ``trivial'' if its underlying space
 is a product of one of its boundaries with the interval (resp. a product
of one of its boundary strata with the interval modulo collapsing
the product of the bounding corner with the interval back onto the corner).
Note that a trivial cobordism or cobordism with corners need not be
the identity cobordism---the attaching maps at the
ends could be different. However, a trivial cobordism is manifestly
invertible.

In the definitions below, the non-categorically
minded reader is advised on first reading to read only the
bold-faced portions of the definitions. These give the essential flavor
of the definition, without going into excessive categorical detail.

\label{3dfact}
A {\bf 3D TQFT with factorization} is a monoidal bifunctor
from $3-cobord_2$ to ${\bf VECT}-mod$.

Less briefly, but more intelligibly
to the non-categorically minded, this {\bf entails an assignment of}

\begin{enumerate}
\item  {\bf A finitely generated
semisimple {\bf C}-linear category to each compact 1-manifold.}
\item An exact {\bf C}-linear functor to each 2-dimensional cobordism. In
particular, since an exact
{\bf C}-linear functor from {\bf VECT} to a semisimple
{\bf C}-linear category is completely determined by the image of {\bf C},
we have {\bf a choice of an object in the category associated to the boundary
of each oriented surface with boundary, and more particularly, we have
an assignment of a vector space to every closed oriented surface.}
\item A natural tranformation to each 3-dimensional cobordism with
corners.  In particular {\bf for a 3-manifold with boundary and corners
consisting of two surfaces with boundary sharing their common
boundary as a corner, we have a map in the
category associated to the boundary of the surfaces between the
objects associated to the surfaces.}  Likewise, since the empty surface
is assigned {\bf C}, {\bf a 3-manifold with boundary is assigned a vector
in the vector space associated to its boundary, and finally
a 3-manifold without boundary is assigned a number.}
\end{enumerate}

Moreover, these assignments will satisfy:

{\bf
\begin{enumerate}
\item  The disjoint union of two 1-manifolds gets the tensor product in the
sense of [21] of
the semisimple categories attached to the parts. The empty 1-manifold
will be assigned {\bf VECT}.
\item The 1-manifold with opposite orientation is assigned the dual
category. (cf. Yetter [22])
\item The disjoint union of surfaces is assigned the tensor product of
the vector spaces on the surfaces.
\item The surface with opposite orientation is assigned the dual vector
space.
\item If we cut an oriented surface along a 1-manifold (union of circles),
the vector space on the closed surface is naturally isomorphic to the
hom set of the two objects in the category corresponding to the cuts
which correspond to the two surfaces with boundary. A similar result
holds for the case when we cut a surface with boundary and take hom
with respect to the ``tensor indices'' corresponding to the cuts only.
\item If we cut a 3-manifold along a surface with boundary, the number
invariant of the manifold is the dual pairing of the vectors
associated to the two manifolds with boundary.
\item If we join two cobordisms with corners to form a cobordism, the
linear map associated to the cobordism is the hom of the two linear
maps. If we join two cobordisms with corners along a surface with
boundary to form a new cobordism with the same corner, the map
corresponding to the new cobordism is the composite of the old maps.
\end{enumerate}
}

The reader will no doubt have noticed that these assignments and
conditions fall into two analogous tiers, with semisimple linear
categories closely paralleling vector spaces. The situation for D=4
will be closely analogous again, with a third categorical tier.

It follows from these axioms that the category on a circle has an
associative tensor product, corresponding to the three holed sphere,
or trinion, with associativity constraints given by trivial cobordisms
with corners. Moreover, the category must be braided, again with structure
maps given by trivial cobordisms with corners.
\bigskip

B. Categorification

\bigskip

As we stated above, factorizable TQFTs have tiers of analogous
structures at higher categorical levels. The structure on surfaces in
a 3D-TQFT, for example, is a categorical analog of the structure of a
2D-TQFT.

To make this notion precise, we have defined the idea of a
categorification [23].
The categorification of any type of algebraic structure is a type of
structure
with analogous operations one categorical level higher. A
categorification of a particular algebraic structure is an example of
the categorified type of the structure which when "traced down" gives
us back the original structure.

Let us elucidate these operations by means of a concrete example. It
is common knowledge that the axioms of a category with direct sum and
tensor product (such as the category of vector spaces of finite
dimension) satisfies a list of axioms like those of a ring {except for
additive inverses), only with circles around the operations. The
equations for the ring become isomorphisms for the category, which
then satisfy certain natural axioms called coherence identities [21].

This is what we mean by an analogous structure one categorical level
up. We say that the structure of a semisimple ring category ( a
category with direct sum and tensor product and the usual axioms,
where every object is a sum of irreducibles and hom sets are vector
spaces) is a categorification of the structure of an algebra.

We trace out a tensor category by using the operations to induce a
ring structure on the space of formal linear combinations of objects
in the category. We use real or complex coefficients. Thus, the
category of vector spaces is a categorification of the algebra of real
or complex numbers. (Every object is a sum of copies of a single
generator).

We have discovered that in many situations if a given type of
algebraic structure can be used to generate a TQFT in D=n then a
categorification can naturally be used to generate a TQFT in D=n+1.
This principle, which we have called the dimensional ladder [23],
suggests that we can categorically lift particular TQFTs if we can
find categorifications of the particular structure used to construct
them.

The algebraic operation of going down in categorical level is referred
to as tracing out, because the structure spaces of the category are
replaced by their dimensions as structure coefficients of the algebra.
(Sometimes spaces are assigned exotic or "quantum" dimensions.)

The algebraic operation of tracing out corresponds to the geometric
operation of taking the cartesian product with $S^1$. We can always
turn an n+1 dimensional TQFT into an n dimensional one by this means,
assigning to every object the algebraic structure assigned to its
product with the circle.

Categorification is analogous in several respects to quantization. In
the first place, it is neither always defined nor unique. In the
second, it is the inverse of a well defined procedure (tracing out)
which yields a
simpler theory from a more complex one. (Like Bohr's correspondence
principle).
Finally, categorification of a structure requires that it have an
integral structure, ie a basis in which the structure coefficients are
positive integers. This is reminiscent of the discrete orbitals of
quantum theory.

The reason for hoping that a categorification for CSW theory exists is
that the quantum groups admit categorifications, as part of the
canonical basis program of Lusztig [24]. In [23], we have proposed a
construction of a 4D-TQFT based upon this fact.

\bigskip

D. Topological State Sums

\bigskip

We have already given an example of what we mean by a topological
state sum (TSS) in the formula we cited for D=3. In general, we say we have
a topological state sum whenever we have rules for labelling
triangulations at certain places (specifically "combinatorial flags",
or places where simplices of different dimensions meet) and rules for
extracting numbers from the combinations of labels on simplices, so
that the sum over labellings of the product over simplices of the
numbers is invariant of the triangulation of the manifold.
Such a sum of products has an obvious analogy to a path integral, so
it is no surprise that a TQFT can be constructed from it.

In fact, interesting examples of 2D and 3D TSSs were constructed by
physicists interested in  lattice QFTs in [25,26]. They found that 2D
theories were related to semisimple algebras, while 3D theories were
related to Hopf algebras. The latter work is equivalent to the work of
Kuperberg in [15], although the notation is very different. This is
another example of the relationship between TQFT and abstract or
categorical algebra.

For the purposes of this work, we are especially interested in the
construction of 4D TSSs. The literature on this is quite thin. Aside
from the construction related to a finite group in [27], which seems far
removed from realistic physics, (although it does reproduce
Dijkgraf-Witten theory), there are only two formulas proposed: the one
in [28], which is closely connected to CSW theory, and also uses a
modular tensor category, and the one in [23], which is part of the
dimensional ladder picture, and utilizes the categorification of a
Hopf algebra, or a Hopf category.

The first of these expressions is known to give the euler character and
signature of a 4-manifold, while the second has not yet been explored.
We do not know at this time if the two formulas are related in any
way.

It would probably not be productive for the purposes of this paper to
reproduce the explanations of these two formulas. Let me give a simple
description of them as background for the discussion of possibilities
for their physical interpretation. The first of the two formulas [28]
(formula 4A) is a rather direct analog of the 3D formula discussed
above. Spins are places on faces in the triangulated 4 manifold, and
simple closed networks are associated to the boundaries of the
4-simplices. The formula is then a sum over labellings of a product
over simplices, as usual. The second formula [23] (formula 4B), is much
more subtle. It uses the structure of a Hopf category associated to
the Borel subalgebra of a quantum group. These exist in general
because of Lusztig's work, but the SU(2) case can be done by hand.

\bigskip

V. TQFT AND QUANTUM GRAVITY

\bigskip

Now let us state a conjecture which summarizes all the aims of the
above mathematical program.

{\bf CONJECTURE:} {\it there exists a 4D-TQFT which is a
categorification of CSW theory, is factorizable, and can be written in
terms of a topological state sum. The topological state sum can be
interpreted as a quantum geometry, and gives Einstein's equations in a
suitable classical steepest descent approximation.}

This is a sizable wish list. In a moment I will review how much of
this is known or highly likely, and how much is plausible.

First, let us see how a theory of quantum gravity could be constructed
along these lines which could be tested by experiment.

We could model an experiment by means of a four manifold
with corners. The initial state could be defined by picking a state
on the boundary surface of the initial manifold with boundary, and a
final state could be similarly defined on the terminal hypersurface
with boundary. We could then calculate probability amplitudes by
computing the state sum on the 4 manifold with corners.
Physically, we would think of this as doing a quantum
experiment on the outside of a classical observer, while
fixing the state of the observer itself. We would see excitations
which propagated, which were localized relative to the observers on
the boundaries, but not more finely than the Planck scale. The fact
that steepest descent, in a classical limit, for the TSS gave us back
Einstein's equation would imply that we had a possible quantization of
general relativity. The observers would solve the problem of locality,
which is as things should be.

Why do we consider it desirable that the 4D TQFT be a categorification
of CSW theory? The argument is the following: if the 4D theory is the
categorification of CSW, then we must reproduce CSW if we trace out
the 4D theory, ie consider it on cartesian products with $S^1$.

The operation of tracing out is well known to relativiste in a
different setting. It is called "quantization with periodic euclidean
time." It isgenerally believed that this procedure produces a thermal
state; much of Hawking's work on black hole radiation is based on this
idea.

Thus, the idea that CSW is recovered by tracing out is consistent with
the suggestion [29] that CSW
is a thermal state for quantum gravity. The suggestion in [30] that the
CSW lagrangian provides a time parameter for quantum gravity is also
closely related.

How likely then, is it that our mathematical conjecture can be
demonstrated? Most of it seems to be within reach. Formula 4B is
directly a quantization of Kuperberg's formula [15], which comes rather
close to reproducing CSW. On the other hand, 4A, although it produces
only a classical invariant, is extremely similar to the CSW
invariants.

The really difficult piece of our wish list is the geometric
interpretation, and the recovery of Einstein's equation from the
steepest descent of the formula in the classical limit. (The classical
limit is the limit of large spins, ie of classical distances.) It
seems likely that if we could produce a state sum which imitated
Einstein's equation in 4D, then even the most categorically disinclined
relativists would agree that we had something of interest.

The motivation for hoping this from the mathematical point of view is,
in the first place, the fact that exactly what we want happens in the
3D case. The evaluation of a spin network reproduces 3D Einstein in
the classical limit, see [13]. The CSW invariant is very similar, and
only adds a cosmological constant. Now the similarities of the 3D and
4D formulas give us reason to hope. Formula 4A can be thought of as
putting spins on surfaces in 4D, which is reminiscent of the recent
suggestion [31] that the observables for the loop variables are areas of
surfaces, not lengths.

In fact, it is possible, up to a small ambiguity, to define the
geometry of a triangulated 4-manifold by assigning areas to faces,
instead of lengths to edges. My collaborators and I are currently
exploring whether this gives rise to an interesting interpretation of
formula 4A.

The suggestion that the quantum theory of gravity can be
reconstructed from a topological state sum, and hence indirectly from
abstract algebra can be thought of several ways. It
is reminiscent of the suggestion of E. Witten that the fundamental
unified theory of nature should resemble a "sporadic" algebraic
structure. Although Witten's suggestion was motivated by mathematical
aspects of particle physics, it has a general philosophical appeal as well.
The picture that some very special algebraic structure, whose
operations are naturally related to four dimensional geometry
creates our world has a
strange appealing simplicity.

\bigskip
VI. ANCESTRAL VOICES

\bigskip

To my great surprise, I recently learned from an essay by John
Stachel [32], that the ideas explored here are very similar to the ones
that Einstein was exploring near the end of his career. Apparently,
these ideas remained obscure because Einstein did not have the
mathematical tools to carry them to fruition, and discussed them only
orally or in letters.

Let us repeat a few of Einstein's sentences:

{\it  ... you have correctly grasped the drawback that the continuum
brings. If the molecular view of matter is the correct (appropriate)
one, ie., if a part of the universe is to be represented by a finite
number of moving points, then the continuum of the present theory
contains too great a manifold of possibilities. I also believe that
this too great is responsible for the fact that our present means of
description miscarry with the quantum theory. The problem seems to me
how one can formulate statements about a discontinuum without calling
on a continuum as an aid; the latter should be banned from the theory
as a supplementary construction not justified by the essence of the
problem, which corresponds to nothing "real". But we still lack the
mathematical structure unfortunately...}

\bigskip

....Einstein to Walter Dallenbach

\bigskip

So not only was Einstein interested in a discrete theory, but he
thought the missing element was mathematical.

\bigskip

{\it The other possibility leads in my opinion to a renunciation of
the space-time continuum, and to a purely algebraic physics.}

\bigskip

...Einstein to Paul Langevin

\bigskip

I am proposing to scrap the metrical continuum, but retain the
topological one. Algebra!

\bigskip

{\it An algebraic theory of physics is affected with just the
inverted strengths and weaknesses; aside from the fact that no one has
been able to propose a possible logical schema for such a theory. It
would be especially difficult to derive something like a
spatio-temporal quasi-order from such a schema. I cannot imagine how
thw axiomatic framework of such a physics would appear, and I don't
like it when one talks about it in dark apostrophies. But I hold it
entirely possible that the development will lead there...}

\bigskip

.....Einstein to H. S. Joachim.

\bigskip

The line of development in this paper also seems relevant to the
objections to quantum mechanics of the {\bf LATE} Einstein. In his
later writings, Einstein was bothered by the fact that quantum
mechanics posits a macroworld, and gets its observables from it.
Since I am proposing a picture in which the classical world reappears
in a suitable limit, I believe that the objection Einstein voiced
could be answered by the completion of my program.

I was able to resist the urge to retitle this paper "On some ideas of
Professor Einstein," but given the unconventional nature of my
proposal, I am happy for the moral support. Einstein, after all, was
often ahead of his time.

\bigskip

Acknowledgements: The ideas of this paper have evolved in the course
of an ongoing dialog of many years with Lee Smolin and Carlo Rovelli.
The mathematics has evolved largely in the course of collaborations
with Igor Frenkel and David Yetter. A few sentences were lifted from a
joint paper with David, and represent joint work. The quotations from
Einstein are from Stachel. Abhay Ashtekar knocked a few of the bumps
off this proposal durring a recent visit to Penn. There are still more
to knock.

\bigskip

BIBLIOGRAPHY

\smallskip

[1]  M. Atiyah
, Topological Quantum Field Theory
, Cambridge University Press
, 1990.

\smallskip

[2]  E. Witten
, Topological quantum field theory
, Comm. Math. Phys.
117 , 353-386, 1988.

\smallskip

[3]  N. Reshetikhin and V.G. Turaev
, Invariants of 3-manifold via link polynomials and quantum
groups
, Invent. math.
103 547-597,  1991.

\smallskip

[4] L. Crane, 2-d Physics and 3-d Topology, Commun. Math. Phys.
135 615-640, 1991.

[5]  V. Turaev and O. Viro
, State Sum Invariants of $3$-manifolds and Quantum $6j$ symbols
, Topology
 31 865-902,  1992

\smallskip

[6] A.Ashtekar, Lectures on Non-Perturbative Canonical Gravity, world
scientific, Singapore, 1991

\smallskip

[7] S. Mac Lane, Categories for the Working Mathematician, Springer-Verlag,
1971.

\smallskip

[8] J. Baez, Quantum gravity and the Akgebra of Tangles, 1992

\smallskip

[9] D.N. Yetter, State-sum invariants of 3-manifolds associated to artinian
semisimple tortile categories, Top. and its App. 58 47-80, 1994.

\smallskip

[10]. G. Moore and N. Seiberg, Classical and Quantum Conformal Field
Theory, Commun Math. Phys. 123 177-254 (1989)

\smallskip

[11] R. Penrose, Angular Momentum, an Approach to Combinatorial
Space-Time, in Quantum Theory and Beyond, T. Bustin, ed.

\smallskip

[13]. G. Ponzano and T. Regge, Semiclassical Limits of Racah Coefficients
in Spectroscopic and Group theoretical methods in Physics. ed F. Bloch
(North Holland, Amsterdam)

\smallskip

[14] L. Crane, Conformal Field Theory, Spin Geometry, and Quantum
Gravity, Phys. Lett. B v.259 no. 3 1991

\smallskip

[15] G. Kuperberg
, Involutory Hopf Algebras and Three-manifold Invariants,
 Int. J. Math. p41-61, 1989

\smallskip

See also : Non-Involutory Hopf algebras and 3-Manifold Invariants,
preprint, U. of Chicago math department

\smallskip

[16] A. P. Yutsis, I. B. Levinson and V. V. Vanagas Mathematical
applications of the Theory of Angular Momentum, Israel Program of
Scientific Translations, Jerusalem, 1962

\smallskip

[17] E. Witten, 2=1 Gravity as an Exactly Soluble System, Nuc. Phys.
B311 p46 1988

\smallskip

[18] B. Brugmann, R. Gambini, and Jorge Pullin Knot Invariants as
Nondegenerate Quantum Geometries PRL v68 no. 4 1992

[19] E. Witten, Quantum Field Theory and the Jones Polynomial, CMP
v121 p357, 1989

\smallskip

[20] L. Crane and D. Yetter, On Algebraic Structures Implicit in
Topological Quantum field Theories, submitted to CMP, Hepth preprint
9412025

\smallskip

[21] M. Kapranov and V. Voevodsky, Braided Monoidal 2-Categories,
2-Vector Spaces and Zamolodchikov's Tetrahedra Equation,
preprint.

\smallskip

[22] D.N. Yetter, Categorical linear algebra---A setting for questions
from physics and low-dimensional topology, preprint, 1993.

\smallskip

[23] L.Crane and Igor B. Frenkel Four Dimensional Topological Quantum
Field Theory, Hopf Categories, and the Canonical Bases J. Math
Physics, special issue, p5136, 1994

\smallskip

See also: L. Crane and Igor B. Frenkel, On The Dimensional Ladder, to  appear.

\smallskip

[24]  G. Lusztig
, Introduction to Quantum Groups
, Birkhauser ,  1993

\smallskip

[25]  M. Fukama, S. Hosong, and H. Kawai
, Lattice Topological Field Theory
, preprint,  1993

\smallskip

[26] S. Chung, M. Fukama, and A. Shapere,
, Structure of Topological Field Theories in Three Dimensions
preprint

\smallskip

[27] D. Yetter Topological Quantum Field Theories associated to Finite
Groups and
Crossed G-Sets, J. Knot Th. Ram. v1 no 1 p1 1992

\smallskip

see also: R. Dijkgraf and E. Witten, Topological Gauge Theories and Group
Cohomology, Commun. Math. Phys 129 393-429, 1990

\smallskip

[28] L. Crane and D. Yetter, A Categorical Construction of 4D-TQFTs,
in Quantum Topology, World Scientific, 1993 L. Kauffman, ed.

\smallskip

[29] Lee Smolin and Chopin Soo, The Chern Simons Invariant as the
Natural Time Variable, submitted to Cl. and Q. Grav.

\smallskip

[30] A. Connes and C. Rovelli, Von Neumann algebra automorphisms and
the time-thermodynamics relation in  generally covariant quantum
theories
to appear Cl. Q. Grav.

\smallskip

[31] C. Rovelli and L. Smolin, Discreteness of Area and
Volume in Quantum Gravity, to appear in Nuc. Phys. B

\smallskip

[32] J. Stachel Einstein and Quantum Mechanics in Conceptual Problems
of Quantum Gravity Birkhauser 1991 A. Ashtekar and J. Stachel eds

\end{document}